\documentclass[12pt]{article}

\usepackage[english]{babel}

\usepackage[letterpaper,top=2cm,bottom=2cm,left=3cm,right=3cm,marginparwidth=1.75cm]{geometry}
\usepackage{booktabs}
\usepackage{longtable}
\usepackage{indentfirst}
\usepackage{amsmath}
\usepackage{graphicx}
\usepackage{float}
\usepackage[colorlinks=true, allcolors=blue]{hyperref}
\usepackage{natbib}
\usepackage{rotating}
\usepackage{caption}
\usepackage{subcaption}
\usepackage{authblk}
\usepackage{setspace}

\title{Non-fungible token transactions: data and challenges}
\author[1]{Jason B. Cho\thanks{bc454@cornell.edu}}
\author[2,3]{Sven Serneels\thanks{sven@higallop.com}}
\author[1]{David S. Matteson\thanks{dm484@cornell.edu}}
\affil[1]{Department of Statistics and Data Science, Cornell University, Ithaca, NY, USA}
\affil[2]{Gallop Data, Inc., Denver, CO, USA}
\affil[3]{Department of Mathematics, University of Antwerp, Antwerp, Belgium}
\begin{document}
\maketitle

\begin{abstract}
Non-fungible tokens (NFT) have recently emerged as a novel blockchain hosted financial asset class that has attracted major transaction volumes. Investment decisions rely on data and adequate preprocessing and application of analytics to them. Both owing to the non-fungible nature of the tokens and to a blockchain being the primary data source, NFT transaction data pose several challenges not commonly encountered in traditional financial data. Using data that consist of the transaction history of eight highly valued NFT collections, a selection of such challenges is illustrated. These are: price differentiation by token traits, the possible existence of lateral swaps and wash trades in the transaction history and finally, severe volatility. While this paper merely scratches the surface of how data analytics can be applied in this context, the data and challenges laid out here may present opportunities for future research on the topic. \bigskip 

\end{abstract}



\section{Introduction}

The blockchain technology has attracted a lot of attention over the past decade as a decentralized ledger. The blockchain technology was initially launched in 2009 as the {\em Bitcoin Network} based on a whitepaper with sole focus on the transfer and storage of digital currency \citep{Satoshi}. Bitcoin's success as a cryptocurrency sparked a wave of development in blockchain technologies in the wake of which many new blockchains were launched. Some of these concerned refining the Bitcoin approach to digital currencies. However, it was soon recognized that the blockchain technology could have utility much broader than digital currency. The Ethereum network was established to fill that void: it was designed to be a blockchain that enabled to settle digital contracts and run decentralized applications \citep{Vitalik}. Several other networks have followed Ethereum's footsteps since. 

Digital, or {\em smart}, contracts are applied to various fields. One such area that stands out is a contract that transfers a digital right of ownership to a unique asset, which can be both digital or physical. From the blockchain perspective, such an asset is a token, yet it is unique and therefore non-fungible. Again, the concept of non-fungible tokens (NFTs) can be applied to many real world applications, ranging from supply chain to digital sports cards and tokens with utility in games, just to name a few. However, NFTs have probably become most hyped in the realm of digital art, which has received significant mainstream media attention over the past four years. The hype around NFTs
has attracted investors individual and institutional, which has caused certain tokens and even entire collections to be very highly valued.

Owing to the sheer amount of money that circulates in the NFT markets, high valued NFTs are widely seen as a novel financial asset class. A broad set of financial services have emerged around NFTs that span lending protocols, hedge and index funds, NFT fractionalization, buy now pay later services and many more. This phenomenon is referred to as the {\em financialization of NFT}. In addition, there has been a significant increase in studies exploring the financial aspects of NFT markets in the recent years (\cite{HongBao}). Studies such as \cite{KongLin}, \cite{Mazur2021-cy}, \cite{Borri2022-ki} and \cite{KO2022102784} analyzed risk-return of NFT in relation to other asset-class such as bond, equity and commodity. All four studies conclude NFT market to be extremely volatile and behave differently from the traditional asset class. 

A few studies also have explored various potential predictors for NFT price. \cite{DOWLING2022102097} and \cite{fintech1030017} explored the correlation between the price of NFT collections and cryptocurrency such as Bitcoin and Ether. Both studies concluded cryptocurrency pricing to be weakly correlated with NFT pricing. The effect of media on NFT price was also explored by \cite{UMAR2022103031} and \cite{kapooretal}. Particularly, \cite{UMAR2022103031} analyzed how media coverage of COVID-19 affects the price and liquidity in the NFT market. \cite{kapooretal} also found features from social media with regards to a NFT collection, such as number of likes, are significant factors in predicting the price of a NFT token. 

Success of financial services and researches on the NFT markets cannot take place without accurate data. The study by \cite{oh_rosen_zhang_2022} claims that the experienced investors outperform inexperienced investors by 10 percentage points per trade, suggesting high degree of informational inefficiency in the NFT market. Data on NFT transactions, however, come with a set of challenges not shared with traditional financial data. These are both the result of the process that generates the data and of the nature of the asset. While studies on NFT often uses such data, challenges and difficulties in analyzing NFT collections data are rarely explored. This paper will highlight these unique challenges by illustrating them based on the transaction history for a set of eight commonly traded, high valued collections of NFTs and will suggest how to pre-process the raw data so that they become amenable to statistical and financial analysis.   

The paper is organized as follows: in Section \ref{sec:Data}, the data themselves are introduced, along with a discussion on specific characteristics of NFT transaction data and some summary statistics. Section \ref{sec:Challenges} introduces a selection of challenges associated with NFT transactional data and some suggestions to pre-process such data appropriately. Eventually, the final Section provides an outlook into the emerging field of analytics for said data.

\section{The Data}\label{sec:Data}
\begin{figure}[h]
\begin{center}    \includegraphics[width=.5\textwidth]{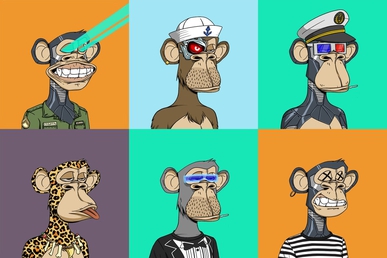}
    \caption{Sample of six {\em Bored Ape Yacht Club} tokens with Blue Beam, Cyborg, 3D (top) and Bored, Holographic, X-Eyes (bottom) Eye traits, among other trait differences.\label{fig:BAYC_token}}
\end{center}
\end{figure}

NFTs have diverse fields of application in the real world. However, the most financialized segments of the broader NFT market are digital art and gaming. Specific aspects of the tokens and their application need to be considered to analyze data for each segment. To narrow down the scope of the paper, the data described here all resort to the digital art segment of {\em profile picture collections (PFP)}. These collections are designed in a similar way: a artist creates a base image, to which attributes are added. By varying these attributes in a way that each realized combination occurs only once, a {\em collection} of unique images is created, each of which follow the base pattern, yet are clearly distinguishable. Ownership for each of these images is then minted on the blockchain of choice as an NFT. An illustration of how varying traits lead to unique images is presented in Figure \ref{fig:BAYC_token} for the collection {\em Bored Ape Yacht Club}. 

\subsection{Sources and Structure}
In theory, all blockchain transactions are publicly available on the blockchain itself. However, querying raw blockchain transactions can be a daunting task and this not only because of the information technical challenges. Transactions that involve one or more smart contracts will be stored on the blockchain as bytecode, which then needs to be decoded and such decoding will be specific to the contract deployed. Because of these technical challenges, several companies have recently spun up data services that provide structred, decoded transactional data for NFTs. One such company is Gallop\footnote{\url{https://www.higallop.com}}, whose set of APIs is the source for the data in this paper. As of this writing, Gallop provides data of 156 Ethereum and 295 Solana NFT collections. The data include transaction histories such as the transacted price of each token, timestamp of the transaction, and the associated wallet addresses, as well as visual traits such as background color, body type, or accessories of the token within the collection. The data selected for this paper correspond to six PFP collections on the Ethereum blockchain (Bored Ape Yacht Club, Bored Ape Kennel Club, Mutant Ape Yacht Club, Cryptopunks, Meebits, and Autoglyphs) and two on the Solana blockchain (Degenerate Ape Academy and Aurory), all of which are highly valued at the time of writing. An overview of the collections selected, is presented in Table \ref{tab:summary}. The data contain transaction histories from collection launch up to March 31, 2022. Data were compiled from the Gallop API responses into a structured table and were analyzed using {\sf R}. Accompanying this manuscript, the tabular data for these eight collections have been made publicly available \citep{DataDOI}. 
\begin{sidewaystable}[p]
\resizebox{\columnwidth}{!}{%
\begin{tabular}{llrrrr}
\hline
\multicolumn{1}{|l}{} &  & \multicolumn{1}{c}{Aurory} & \multicolumn{1}{c}{Autoglyphs} & \multicolumn{1}{c}{Bored Ape Kennel Club} & \multicolumn{1}{c|}{Bored Ape Yacht Club} \\ \hline
\multicolumn{2}{|l|}{Currency} & Solana & Ethereum & Ethereum & \multicolumn{1}{r|}{Ethereum} \\
\multicolumn{2}{|l|}{Project Launch} & August 2021 & April 2019 & June 2021 & \multicolumn{1}{r|}{April 2021} \\
\multicolumn{2}{|l|}{Token Count} & 10,000 & 512 & 10,000 & \multicolumn{1}{r|}{10,000} \\
\multicolumn{2}{|l|}{Transactions} &  &  &  & \multicolumn{1}{r|}{} \\
\multicolumn{1}{|l}{} & \multicolumn{1}{l|}{Daily Average} & 68.78 & 0.45 & 70.54 & \multicolumn{1}{r|}{79.85} \\
\multicolumn{1}{|l}{} & \multicolumn{1}{l|}{Total} & 14,582 & 490 & 20,176 & \multicolumn{1}{r|}{26,751} \\
\multicolumn{2}{|l|}{Floor Price} &  &  &  & \multicolumn{1}{r|}{} \\
\multicolumn{1}{|l}{} & \multicolumn{1}{l|}{Daily Minimum} & 0.005478 & 0.005300 & 0.000010 & \multicolumn{1}{r|}{0.000050} \\
\multicolumn{1}{|l}{} & \multicolumn{1}{l|}{Daily Average} & 21.29 & 51.81 & 1.80 & \multicolumn{1}{r|}{5.72} \\
\multicolumn{1}{|l}{} & \multicolumn{1}{l|}{Daily Maximum} & 35.01 & 449.00 & 8.40 & \multicolumn{1}{r|}{102.69} \\
\multicolumn{2}{|l|}{Total Trading Volume} &  &  &  & \multicolumn{1}{r|}{} \\
\multicolumn{1}{|l}{} & \multicolumn{1}{l|}{Daily Average} & 3,032.63 & 17.95 & 303.64 & \multicolumn{1}{r|}{1,654.62} \\
\multicolumn{1}{|l}{} & \multicolumn{1}{l|}{Total} & 642,917.79 & 19,496.99 & 86,841.95 & \multicolumn{1}{r|}{554,296.23} \\ \hline
 &  & \multicolumn{1}{l}{} & \multicolumn{1}{l}{} & \multicolumn{1}{l}{} & \multicolumn{1}{l}{} \\ \hline
\multicolumn{1}{|l}{} &  & \multicolumn{1}{c}{Cryptopunks} & \multicolumn{1}{c}{Degenerate Ape Academy} & \multicolumn{1}{c}{Meebits} & \multicolumn{1}{c|}{Mutant Ape Yacht Club} \\ \hline
\multicolumn{2}{|l|}{Currency} & Ethereum & Solana & Ethereum & \multicolumn{1}{r|}{Ethereum} \\
\multicolumn{2}{|l|}{Project Launch} & June 2017 & May 2021 & August 2021 & \multicolumn{1}{r|}{August 2021} \\
\multicolumn{2}{|l|}{Token Count} & 9,999 & 10,000 & 20,000 & \multicolumn{1}{r|}{19,414} \\
\multicolumn{2}{|l|}{Transactions} &  &  &  & \multicolumn{1}{r|}{} \\
\multicolumn{1}{|l}{} & \multicolumn{1}{l|}{Daily Average} & 10.86 & 85.60 & 77.01 & \multicolumn{1}{r|}{139.87} \\
\multicolumn{1}{|l}{} & \multicolumn{1}{l|}{Total} & 18,923 & 19,516 & 25,566 & \multicolumn{1}{r|}{30,073} \\
\multicolumn{2}{|l|}{Floor Price} &  &  &  & \multicolumn{1}{r|}{} \\
\multicolumn{1}{|l}{} & \multicolumn{1}{l|}{Daily Minimum} & 0.000050 & 0.052039 & 0.000900 & \multicolumn{1}{r|}{0.000025} \\
\multicolumn{1}{|l}{} & \multicolumn{1}{l|}{Daily Average} & 15.59 & 42.60 & 2.26 & \multicolumn{1}{r|}{0.76} \\
\multicolumn{1}{|l}{} & \multicolumn{1}{l|}{Daily Maximum} & 217.00 & 88.01 & 6.15 & \multicolumn{1}{r|}{15.01} \\
\multicolumn{2}{|l|}{Total Trading Volume} &  &  &  & \multicolumn{1}{r|}{} \\
\multicolumn{1}{|l}{} & \multicolumn{1}{l|}{Daily Average} & 428.46 & 5,222.89 & 9,131.63 & \multicolumn{1}{r|}{1,511.11} \\
\multicolumn{1}{|l}{} & \multicolumn{1}{l|}{Total} & 746,379.75 & 1,190,818.21 & 3,031,702.79 & \multicolumn{1}{r|}{324,889.44} \\ \hline
\end{tabular}%
}
\caption{Associated blockchain, project launch date, token counts, transaction summary, floor price, and trading volume of the eight profile picture token collections}
\label{tab:summary}
\end{sidewaystable}

\subsection{Exploratory Data Analysis}
\begin{figure}[h]
    \centering 
    \begin{subfigure}[b]{0.4\textwidth}
      \includegraphics[width=\linewidth]{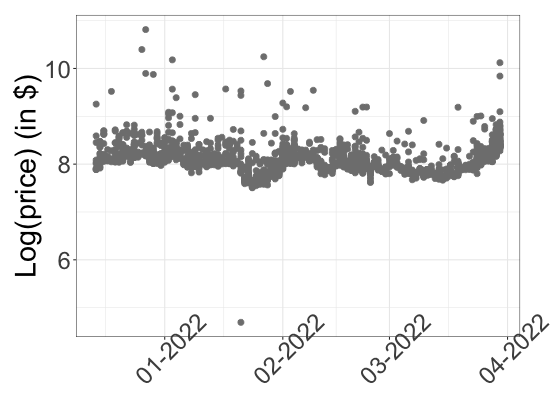}
       \caption{Aurory}
      \label{fig:2_1}
    \end{subfigure}\hfil 
    \begin{subfigure}[b]{0.4\textwidth}
      \includegraphics[width=\linewidth]{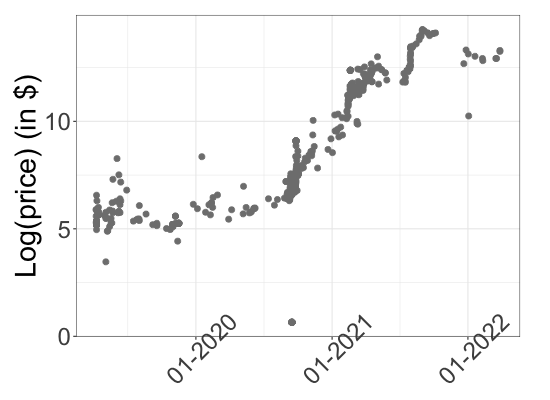}
      \caption{Autoglyphs}
      \label{fig:2_2}
    \end{subfigure}\hfil 
    \medskip
    \begin{subfigure}[b]{0.4\textwidth}
      \includegraphics[width=\linewidth]{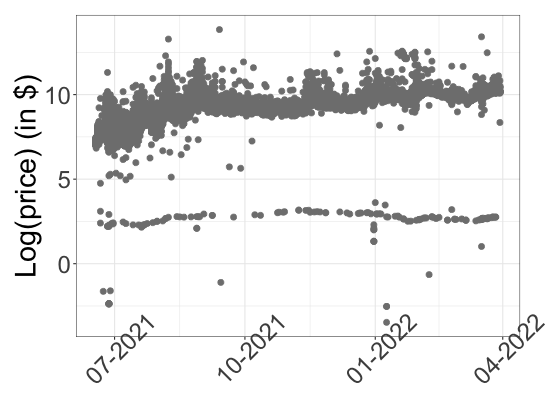}
      \caption{Bored Ape Kennel Club}
      \label{fig:2_3}
    \end{subfigure}\hfil 
    \begin{subfigure}[b]{0.4\textwidth}
      \includegraphics[width=\linewidth]{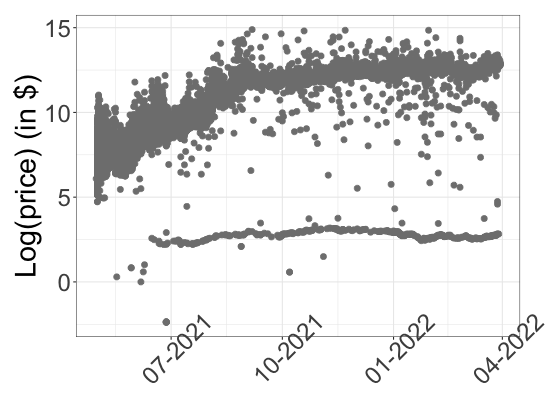}
      \caption{Bored Ape Yacht Club}
      \label{fig:2_4}
    \end{subfigure}\hfil 
    \medskip
    \begin{subfigure}[b]{0.4\textwidth}
      \includegraphics[width=\linewidth]{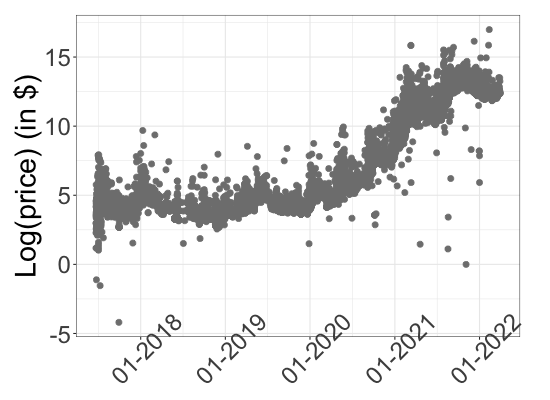}
      \caption{Cryptopunks}
      \label{fig:2_5}
    \end{subfigure}\hfil 
    \begin{subfigure}[b]{0.4\textwidth}
      \includegraphics[width=\linewidth]{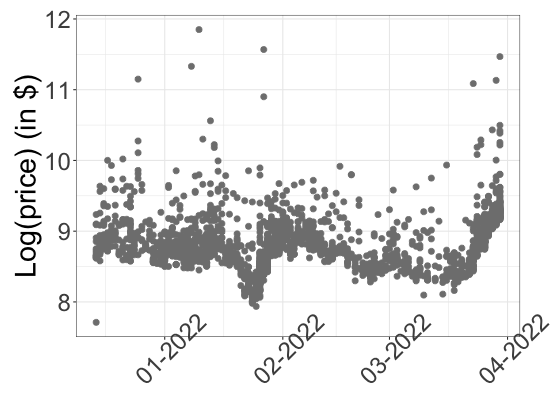}
      \caption{Degenerate Ape Academy}
      \label{fig:2_6}
    \end{subfigure}\hfil 
    \medskip
    \begin{subfigure}[b]{0.4\textwidth}
      \includegraphics[width=\linewidth]{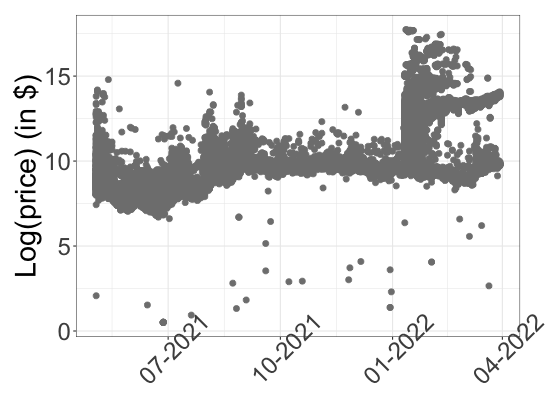}
      \caption{Meebits}
      \label{fig:2_7}
    \end{subfigure}\hfil 
    \begin{subfigure}[b]{0.4\textwidth}
      \includegraphics[width=\linewidth]{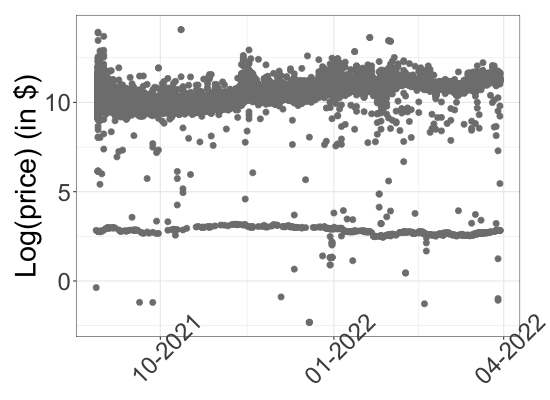}
      \caption{Mutant Ape Yacht Club}
      \label{fig:2_8}
    \end{subfigure} 
    \caption{Log(price) of Eight Profile Picture Collections (PFP) in USD from the Project Launch and March 31, 2022.} \label{fig:Price}
\end{figure}

The average price of a collection, as well as its trend within a given time span, can vary greatly among collections. Price histories are plotted for the eight collections on a logarithmic scale in Figure  \ref{fig:Price}. The average prices of Aurory, Autoglphys, Bored Ape Kennel Club, Bored Ape Yacht Club, Cryptopunks, Degenerate Ape Academy, Meebits, and Mutant Ape Yacht club were 3582.05, 97096.10, 12832.39, 66391.96, 113939.13, 7465.30, 341651.02, and 35896.81 USD, respectively. Only Autoglyphs and Cryptopunks show a steady increase in price. Bored Ape Yacht Club had a steady increase during the first few months after collection launch, but became approximately stationary around August 2021. However, the Aurory, Bored Ape Kennel Club and Mutant Ape Yacht Club collections were stationary during the given period, with the exception of a few outliers. Moreover, it is interesting to note that seemingly bimodal price distributions exist for several collections, most notoriously so for Meebits, an effect that will be discussed in more detail in Sections \ref{sec:latswap} and \ref{sec:wash}.

\begin{figure}[h]
    \centering 
    \begin{subfigure}[b]{0.4\textwidth}
      \includegraphics[width=\linewidth]{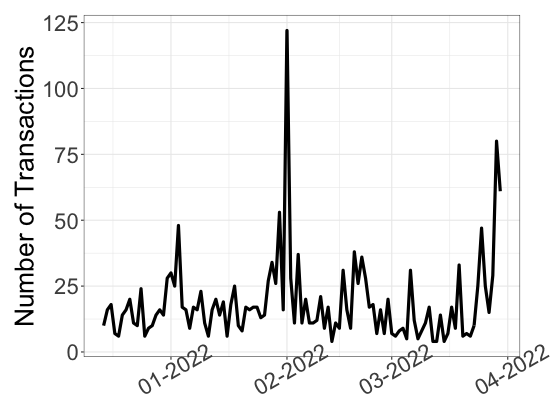}
      \caption{Aurory}
      \label{fig:3_1}
    \end{subfigure}\hfil 
    \begin{subfigure}[b]{0.4\textwidth}
      \includegraphics[width=\linewidth]{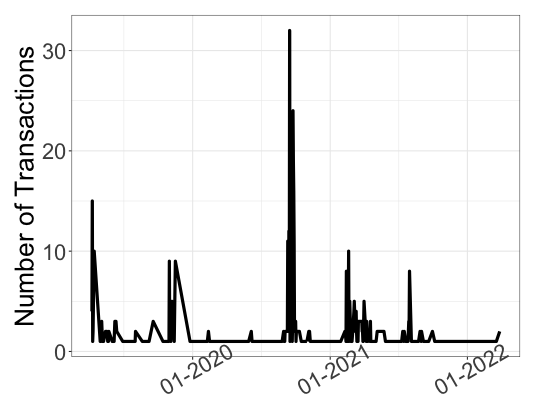}
      \caption{Autoglyphs}
      \label{fig:3_2}
    \end{subfigure}\hfil 
    \medskip
    \begin{subfigure}[b]{0.4\textwidth}
      \includegraphics[width=\linewidth]{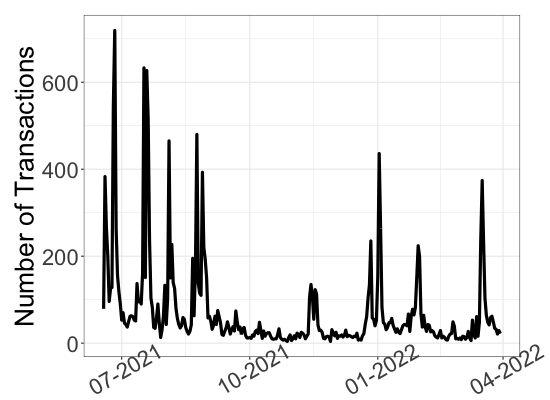}
      \caption{Bored Ape Kennel Club}
      \label{fig:3_3}
    \end{subfigure}\hfil 
    \begin{subfigure}[b]{0.4\textwidth}
      \includegraphics[width=\linewidth]{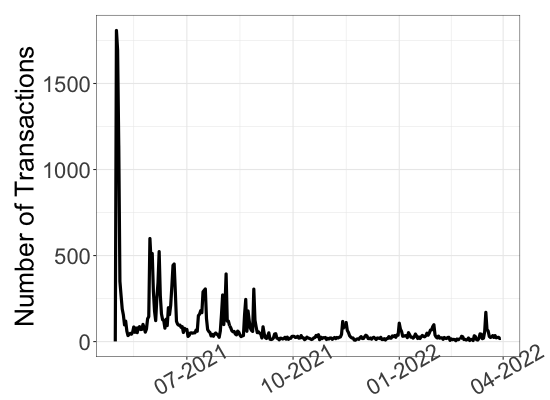}
      \caption{Bored Ape Yacht Club}
      \label{fig:3_4}
    \end{subfigure}\hfil 
    \medskip
    \begin{subfigure}[b]{0.4\textwidth}
      \includegraphics[width=\linewidth]{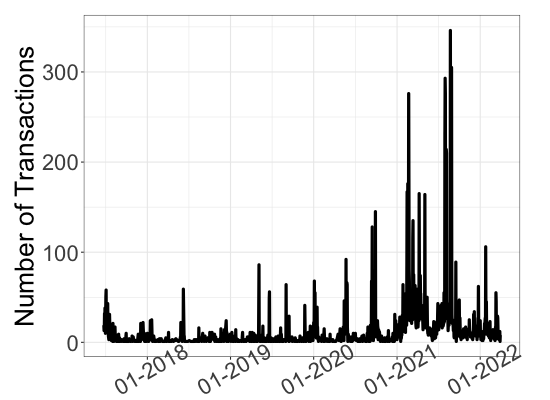}
      \caption{Cryptopunks}
      \label{fig:3_5}
    \end{subfigure}\hfil 
    \begin{subfigure}[b]{0.4\textwidth}
      \includegraphics[width=\linewidth]{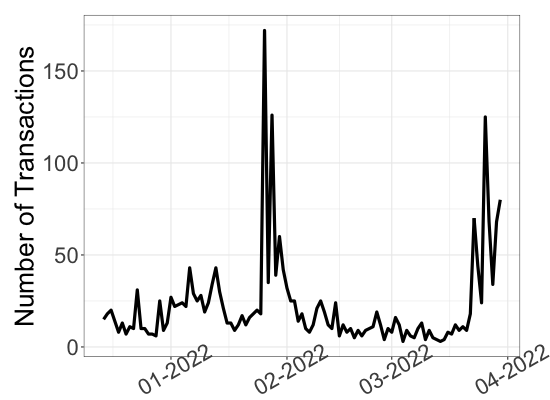}
      \caption{Degenerate Ape Academy}
      \label{fig:3_6}
    \end{subfigure}\hfil 
    \medskip
    \begin{subfigure}[b]{0.4\textwidth}
      \includegraphics[width=\linewidth]{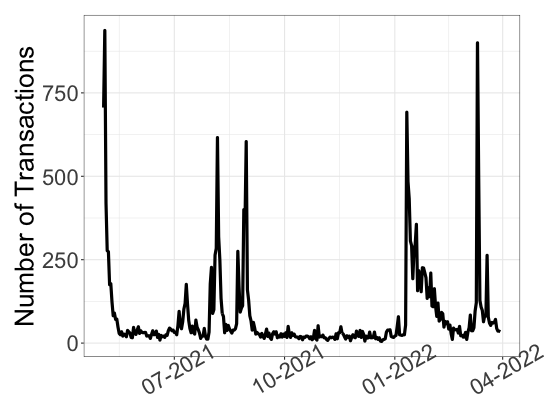}
      \caption{Meebits}
      \label{fig:3_7}
    \end{subfigure}\hfil 
    \begin{subfigure}[b]{0.4\textwidth}
      \includegraphics[width=\linewidth]{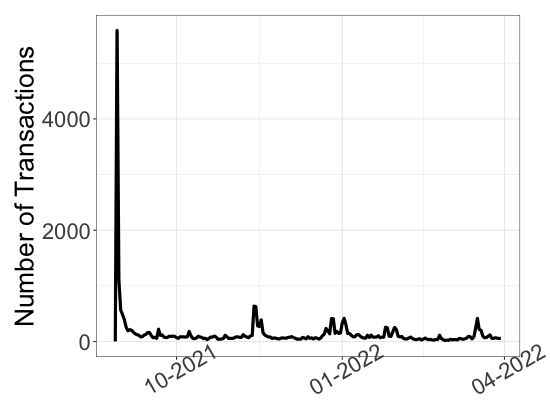}
      \caption{Mutant Ape Yacht Club}
      \label{fig:3_8}
    \end{subfigure} 
    \caption{Number of Transactions of Eight Profile Picture Collections (PFP) in USD from the Project Launch and March 31, 2022.} \label{fig:Transactions}
\end{figure}

In addition to the price, trading volume of a financial asset is another technical indicator that traders look to determine liquidity when making trading decisions. Similar to high trading volume of a stock after the Initial Public Offering (IPO), collections such as Bored Ape Kennel Club, Bored Ape Yacht Club, Meebits, and Mutant Ape Yacht Club hit their highest trading volumes right after project launch as shown in Figure~\ref{fig:Transactions}. When compared against traditional financial instruments, the trading volumes for all eight collections were low. While some collections reached a thousand to few thousand daily transactions, the median daily transactions of Aurory, Autoglphys, Bored Ape Kennel Club, Bored Ape Yacht Club, Cryptopunks, Degenerate Ape Academy, Meebits, and Mutant Ape Yacht club were 16, 1, 36, 33, 5, 13, 33 and 80 respectively. Trading volume also had high day-to-day variation with the standard deviation of 15.99, 3.80, 104.63, 167.80, 28.60, 25.44, 123.02, and 392.53 respectively.

\section{Challenges and Limitations}\label{sec:Challenges}
\subsection{Traits}\label{sec:Traits}

\begin{figure}[h]
    \centering 
    \begin{subfigure}[b]{0.48\textwidth}      \includegraphics[width=\linewidth]{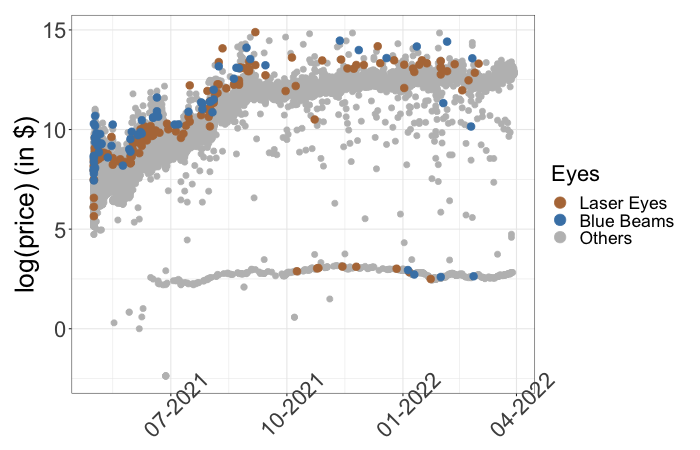}
      \caption{Price Change by Eye Type (Bored Ape Yacht Club)}
      \label{fig:4_1}
    \end{subfigure}\hfil 
    \begin{subfigure}[b]{0.48\textwidth}
      \includegraphics[width=\linewidth]{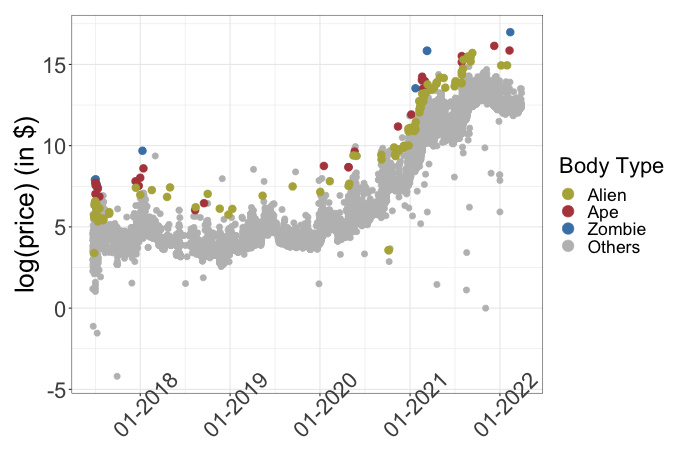}
      \caption{Price Change by Body Type (Cryptopunks)}
      \label{fig:4_2}
    \end{subfigure}\hfil 
    \begin{subfigure}[b]{0.48\textwidth}
      \includegraphics[width=\linewidth]{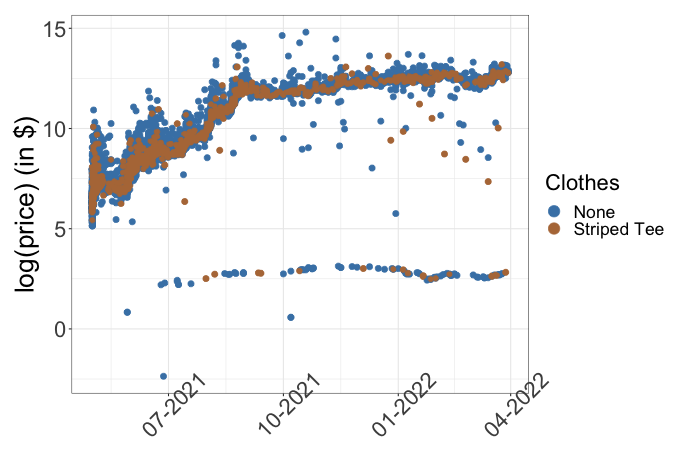}
      \caption{"None" vs "Striped Tee" (Bored Ape Yacht Club)}
      \label{fig:4_3}
    \end{subfigure}
    \caption{Log(price) of Bored Ape Yacht Club and Cryptopunks by Various Traits.}
    \label{fig:rarity}
\end{figure}

One of the main challenges to analyze the price of NFTs at the collection level is the wide price range among tokens, as shown for two collections in Figure~\ref{fig:Price}. Such differences can partially be explained by the design of each token. Tokens are generated according to a base pattern that is then complemented by giving it certain traits. Most collections are designed in such a way that a certain (combination of) traits is much rarer than other combinations. For instance, the Cryptopunks collection consists of 10,000 tokens, distributed across six body types. However, there are only 6 Aliens, 14 Apes and 24 Zombies, while the remainder of the tokens have either Male or Female body type. The price of a token with Alien, Ape and Zombie body types was generally considerably higher than the ones with Male and Female body types (Figure \ref{fig:4_2}). Similarly, for the Bored Ape Yacht Club (BAYC) by YugaLabs, only 118 of 10,000 tokens have laser eyes or blue beams radiating from their eyes, and these were traded at prices noticeably higher than tokens with others Eyes traits (Figure \ref{fig:4_1}).

However, not all design traits seem to contribute to the price of a token. For example, in the BAYC collection, the most common Cloth trait is "none": 1,886 of the 10,000 tokens have this trait. The next common trait is "Striped Tee", which 412 tokens have. While the tokens with "None" trait are about 4.5 times more common than the ones with "Striped Tee" trait, no noticeable differences were found in their prices, as is illustrated in Figure~\ref{fig:4_3}. 

\begin{figure}[h]
\begin{center}
\includegraphics[width=0.5\textwidth]{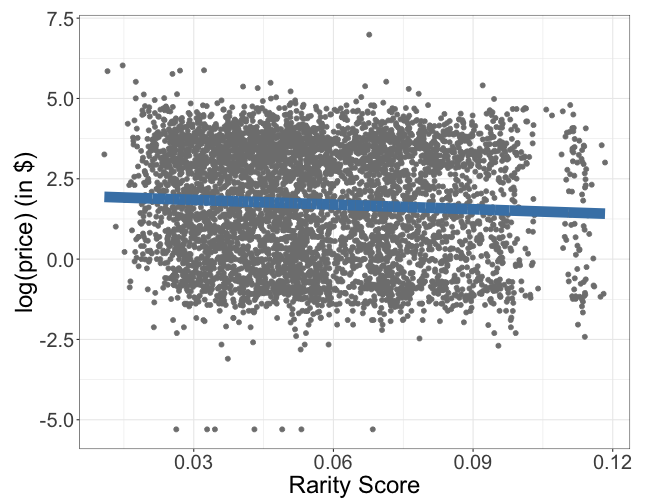}
\caption{Price vs Rarity of a Token (Bored Ape Yacht Club). \label{fig:price_vs_rarity}}
\end{center}
\end{figure}

The relationship between the price of a token and its rarity was explored by calculating the rarity score and analyzing the linear relationship between the score and the average price of the token. The rarity score of a token was calculated by averaging the rarity of a token over multiple traits (hair, background, accessories, etc.) a token is designed to have, where rarity is simply the fraction of the number of tokens with the trait to the total number of tokens in the collection. A negative correlation between rarity and price of a token was detected (Figure~\ref{fig:price_vs_rarity}), suggesting that rarer tokens are indeed traded at higher prices than more common tokens. However, the R-squared of only around 0.001 suggests that the correlation is rather weak, and that there are other meaningful factors that contribute to the price of a token. Here we also note that average rarity is only one way to compute token rarity\footnote{The company \url{https://rarity.tools} specializes in token rarity and offers two version of highly customizable rarity models, the details of which would range beyond the scope of this paper.} and other measures may lead to more significant correlations. For instance, \cite{KongLin} arrive at similar findings using a more sophisticated hedonic regression model for Cryptopunks transactions though May 2021.

\subsection{Lateral swaps}\label{sec:latswap}
\begin{figure}[h]
    \centering 
    \begin{subfigure}[b]{0.5\textwidth}
      \includegraphics[width=\linewidth]{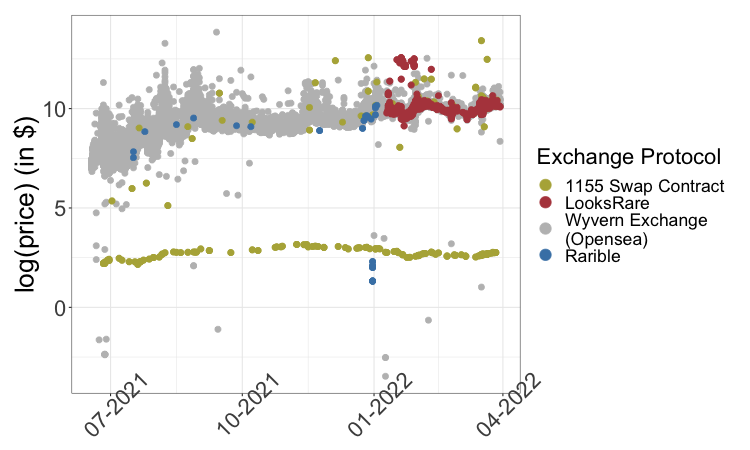}
      \caption[Price Change by Exchange Type (Bored Ape Kennel Club)]
    {\tabular[t]{@{}l@{}}Price Change by Exchange Type \\ (Bored Ape Kennel Club)\endtabular}
      \label{fig:6_1}
    \end{subfigure}\hfil 
    \begin{subfigure}[b]{0.5\textwidth}
      \includegraphics[width=\linewidth]{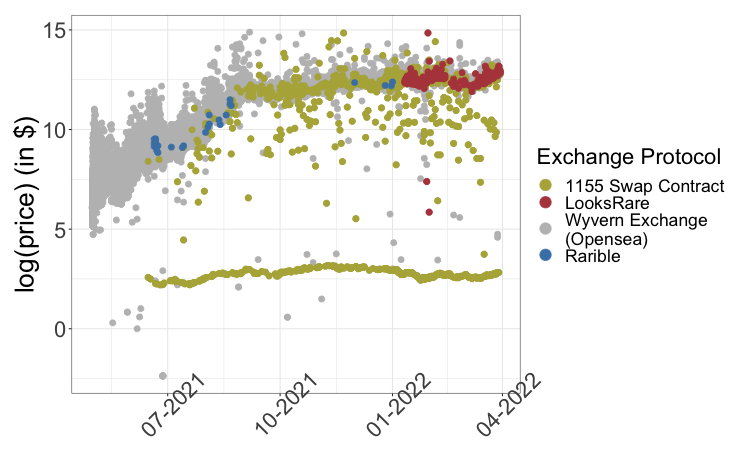}
      \caption[Price Change by Exchange Type (Bored Ape Yacht Club)]
    {\tabular[t]{@{}l@{}}Price Change by Exchange Type \\ (Bored Ape Yacht Club)\endtabular}
      \label{fig:6_2}
    \end{subfigure}\hfil 
    \begin{subfigure}[b]{0.5\textwidth}
      \includegraphics[width=\linewidth]{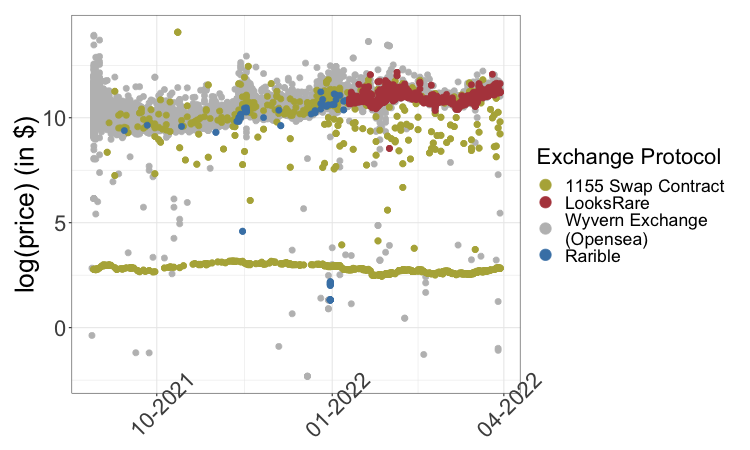}
      \caption[Price Change by Exchange Type (Mutant Ape Yacht Club)]
    {\tabular[t]{@{}l@{}}Price Change by Exchange Type \\ (Mutant Ape Yacht Club)\endtabular}
      \label{fig:6_3}
    \end{subfigure}\hfil 
    \begin{subfigure}[b]{0.5\textwidth}
      \includegraphics[width=\linewidth]{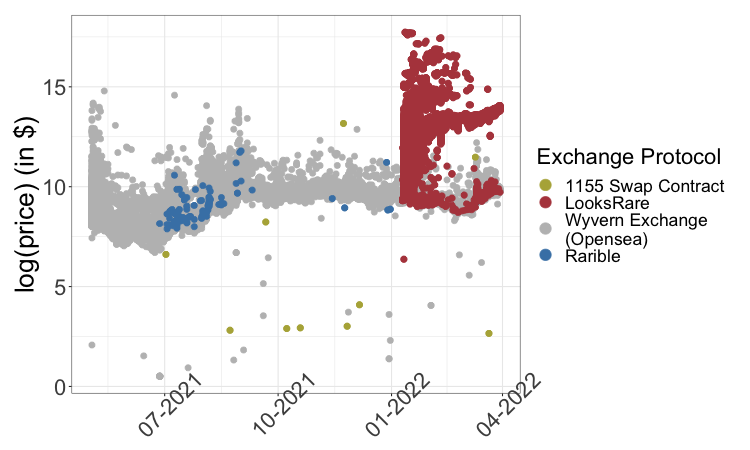}
      \caption[Price Change by Exchange Type (Meebits)]
    {\tabular[t]{@{}l@{}}Price Change by Exchange Type \\ (Meebits)\endtabular}
      \label{fig:6_4}
    \end{subfigure}
    \caption{Price Change by Exchange Type}
    \label{fig:exchange_type}
\end{figure}

One challenge to decode NFT transactions on Ethereum resides in the fact that while transactions have a \texttt{value} field, the value contained in that field reflects the value paid into the top level contract, which will only correspond to the price paid for the token in very simple transactions, such as peer-to-peer Cryptopunk transactions. In transactions that involve more than one smart contract, the transaction price can only be retrieved by decoding further layers of smart contract execution. In fact, many transactions will have a value equal to zero at the top level, but contain non-zero transaction prices in the token contract execution. 

All of the above is accounted for in Gallop's data aggregation, yet some atypical transactions may still be found in decoded transaction data. For instance, when users use a smart contract to swap tokens laterally between wallet addresses, they may still have to pay the execution price of the swap contract, which will appear on the blockchain as the value. This results in a set of transactions recorded at a price of 0.005 ETH, which is considerably lower than the fair value token transaction price, as strikes the eye for collections such as Bored Ape Kennel Club, Bored Ape Yacht Club, and Mutant Ape Yacht Club, plotted in Figure \ref{fig:exchange_type}. As these transaction records are not indicative of the fair value of the token, they should be excluded when performing token or collection valuation analysis\footnote{The public Gallop API will filter these anomalous transactions out, yet they were retained in the data described here as they represent a challenge common in NFT transaction data.}.

\subsection{Wash Trading}\label{sec:wash}

{\em Wash trading} is a process in which a market participant sells and buys the same token back multiple times within a short period to deceive other market participants about its price and liquidity. If such malicious market participants are successful, they can completely distort transaction price statistics. A good illustration of that effect is shown in the transaction price history for Meebits (Figure \ref{fig:6_4}). Meebits' transaction history follows an approximately smooth and continuous trend through January 2022, but then a vast amount of transactions occur at a wide range of prices deviating from that trend. Unusual transactions of the Meebits tokens garnered media attention and a few articles have been published on this phenomenon (\cite{howcroft_2022}). Many suspect that most, if not all, of the transactions at significantly higher prices compared to the trend to be wash trades. As can be seen from Figure~\ref{fig:exchange_type}, which groups transactions by exchange protocol, almost all transactions at elevated prices have been executed using the {\em LooksRare} protocol. 

LooksRare\footnote{\url{https://looksrare.org/}} launched in January 2022 as a new decentralized marketplace, aiming to take on competition with the (then and now) dominant marketplace OpenSea\footnote{\url{https://opensea.io/}}. In order to grow market share, the LooksRare protocol rewards users by paying them back a fraction of the transacted amount in their native LOOKS token. However, the LooksRare protocol thereby inherently encourages wash trading. By using LooksRare, wash traders reap a double benefit: they not only succeed at artificially raising average sales prices and volumes for their tokens, but they also get rewarded in LOOKS token for doing so. Moreover, they can then decide to stake the LOOKS token, for which they will reap a second round of rewards in the form of Wrapped Ether (WETH) tokens.  An example of how trading a single token in this way can lead to profits of approximately 3.5 million USD, is described in \cite{CoffmanBlog}.    

\begin{figure}[h]
    \centering 
    \begin{subfigure}[b]{0.5\textwidth}
      \includegraphics[width=\linewidth]{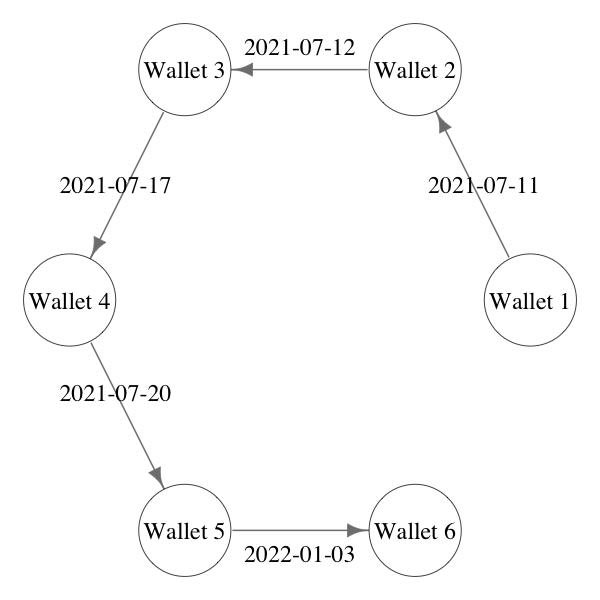}
      \caption[Transactions between 2021-07-11 and 2022-03-07 (Bored Ape Kennel Club \#282)]
      {\tabular[t]{@{}l@{}} Transactions between \\ 2021-07-11 and 2022-01-03 \\ (Bored Ape Kennel Club \#282)\endtabular}
    \label{fig:7_1}
    \end{subfigure}\hfil 
    \begin{subfigure}[b]{0.5\textwidth}
      \includegraphics[width=\linewidth]{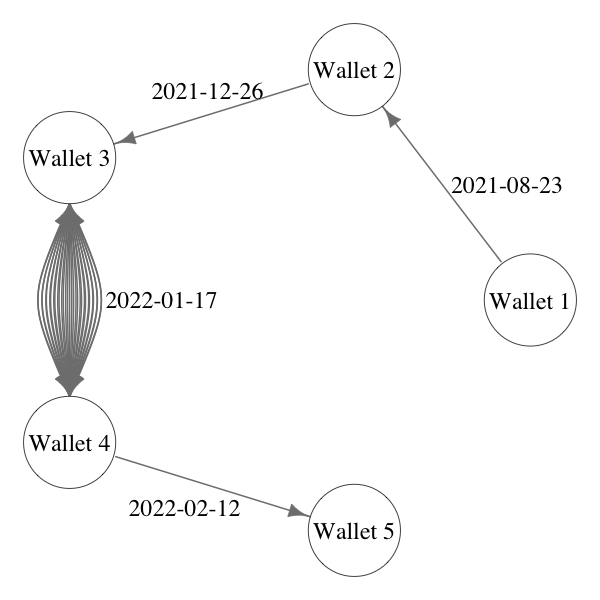}
      \caption[Transactions between 2021-08-23 and 2022-02-12 (Meebit \#17021)]
      {\tabular[t]{@{}l@{}} Transactions between\\ 2021-07-11 and 2022-02-12 \\ (Meebit \#17021)\endtabular}
      \label{fig:7_2}
    \end{subfigure}\hfil 
    \caption{Transactions History of Bored Ape Kennel Club \#282 and Meebits \#17021}
    \label{fig:Token_wallet_transaction}
\end{figure}

While wash trading can clearly distort market prices, it is an entire different question how it can be detected. To propose a detailed algorithm for that purpose would lead us beyond the scope of this Data Note, yet we will provide a hint as to how either wallets or collections can be flagged for suspicious wash trading activity.  

At first, a wash trade is by definition a a trade executed between different wallets owned by the same person or entity. On the blockchain, wallets are anonymous and in theory, users can create new wallet addresses each time they trade. However, at least for non-automated traders or less sophisticated wash trading algorithms, to create a new wallet address each time is a cumbersome step. Therefore, one can expect that in such cases, the token traded would {\em return} to the originating wallet address at a given point in time. In several cases, the latter occurs a striking amount of times, as illustrated in Figure \ref{fig:7_2}. Herein, the trade graph is plotted for two example tokens. On the one hand, Bored Ape Kennel Club \#282 has unique wallets involved in all of its five transactions (Figure \ref{fig:7_1}), whereas Meebit \#17021 was transacted 32 times on January 17, 2022, between the same pair of wallets , notably via LooksRare, before eventually being transacted on Opensea into a different wallet on February 12, 2022. 

\begin{figure}[h]
    \centering 
    \begin{subfigure}[b]{0.4\textwidth}
      \includegraphics[width=\linewidth]{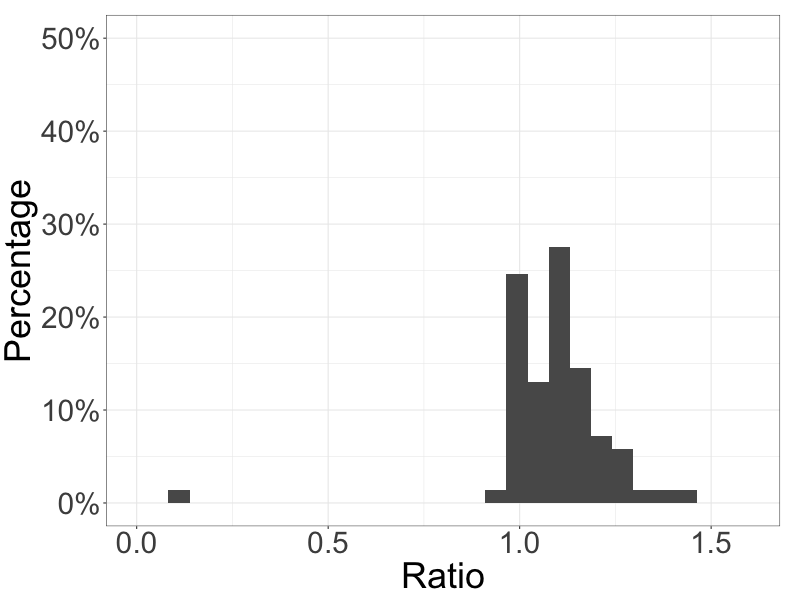}
      \caption{Bored Ape Kennel Club}
      \label{fig:8_1}
    \end{subfigure}\hfil 
    \begin{subfigure}[b]{0.4\textwidth}
      \includegraphics[width=\linewidth]{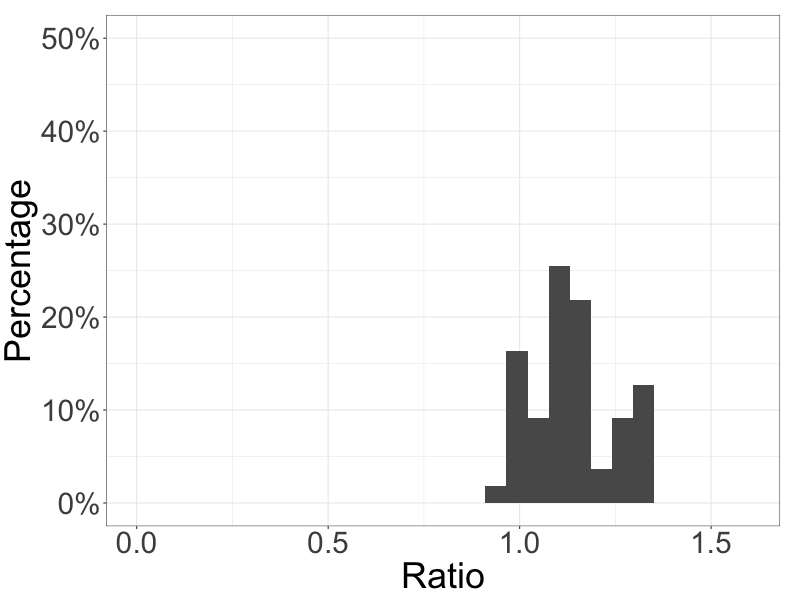}
      \caption{Bored Ape Yacht Club}
      \label{fig:8_2}
    \end{subfigure}\hfil 
    \medskip
    \begin{subfigure}[b]{0.4\textwidth}
      \includegraphics[width=\linewidth]{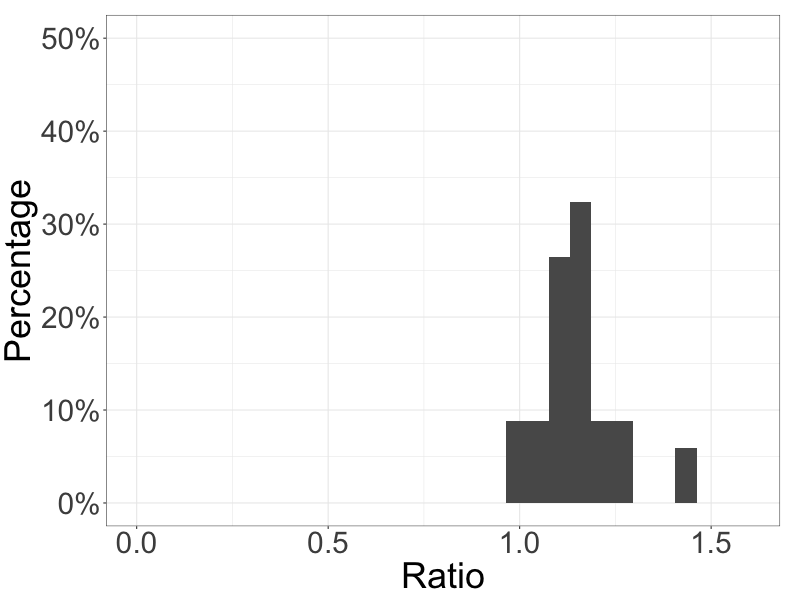}
      \caption{Cryptopunks}
      \label{fig:8_3}
    \end{subfigure}\hfil 
    \begin{subfigure}[b]{0.4\textwidth}
      \includegraphics[width=\linewidth]{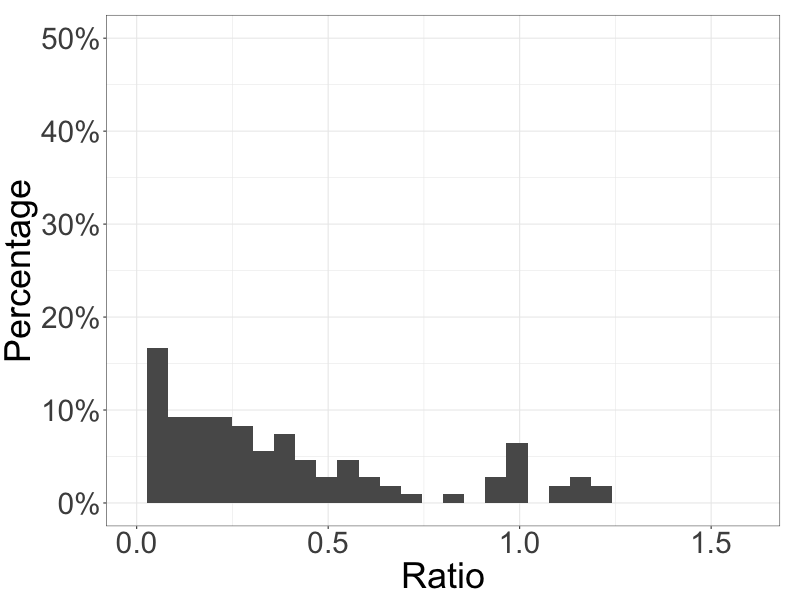}
      \caption{Meebits}
      \label{fig:8_4}
    \end{subfigure}\hfil
    \medskip
    \begin{subfigure}[b]{0.4\textwidth}
      \includegraphics[width=\linewidth]{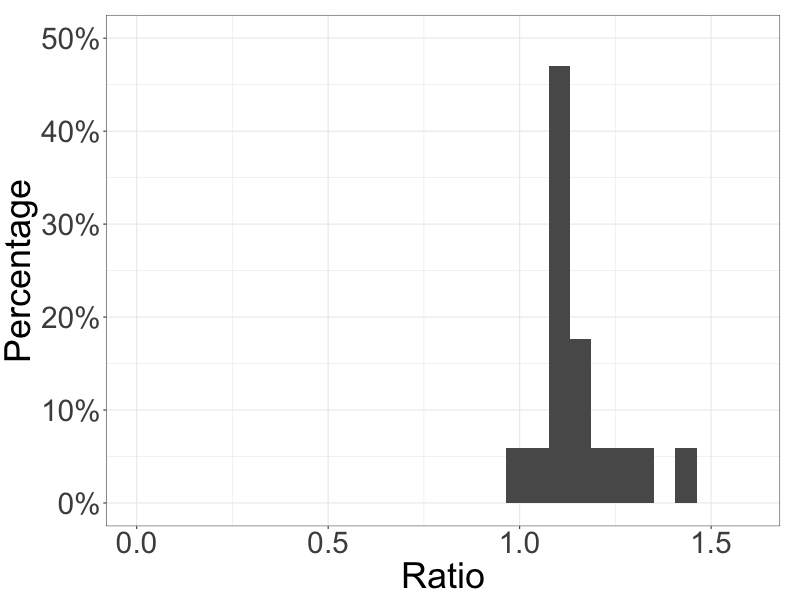}
      \caption{Mutant Ape Yacht Club}
      \label{fig:8_5}
    \end{subfigure}\hfil
    \caption{Distribution of the Wallet to Transactions Ratio.}
    \label{fig:Wallet_Transaction}
\end{figure}

By the anonymous nature of NFT wallets, one can not detect wash trades with absolute certainty. Even though transactions involving BAKC \#282 involve unique wallet addresses, it is possible that all these wallets are owned by the same party. Nevertheless, the 32 transactions between a pair of wallets involving Meebit \#17021 are most likely the result of wash trading activity. Many other examples of repeated transfers between wallet pairs can be found in the Meebits transaction history. The latter would lead us to believe that a lot of the wash trading activity has been executed with a low degree of automation. Two holistic approaches investigating wash trading for the entire blockchain arrive at similar conclusions. At first, \cite{TariqSifat} corroborate the existence of wash trades on the Ethereum and Wax blockchains by performing hypothesis tests on the entirety of NFT transactions. They test for adherence to Benford's law, existence of clusters in pricing and test if transaction prices fit Pareto---Levy's law, all of which significantly point to wash trades being present. Secondly, \cite{vonWaechter} store the entirety of Ethereum transactions that involve NFTs as a directed graph and then extract all closed loops from it, which leads them to conclude that approximately two percent of all NFT transactions are suspect to being wash trades and that the vast majority of these only involve up to a handful of wallet addresses. To identify such closed loop trades, they employ \cite{Johnson}'s exhaustive algorithm to find elementary circuits in a directed graph.   

While \cite{vonWaechter} state to observe suspected wash trading in most collections, some collections have been prone to a higher degree of suspect activity. Figure \ref{fig:Wallet_Transaction} shows the ratio of the number of unique wallets to the total number of transactions within the data history. A low unique wallet-to-transaction ratio would indicate a large number of transactions occurring between only a few wallet addresses, which is indicative of suspected wash trading. Meebits stands out as a collection with a comparably low unique wallets to transaction ratio (Figure~\ref{fig:8_4}). Despite LooksRare's effort to discourage wash trading by raising transaction fees and limiting the amount of LOOKS token to be rewarded in a day, the data suggest a that many continue to exploit the system to this day. 

\subsection{Market Volatility}\label{sec:Volatility}

\begin{figure}[h]
    \centering 
    \begin{subfigure}[b]{0.4\textwidth}
      \includegraphics[width=\linewidth]{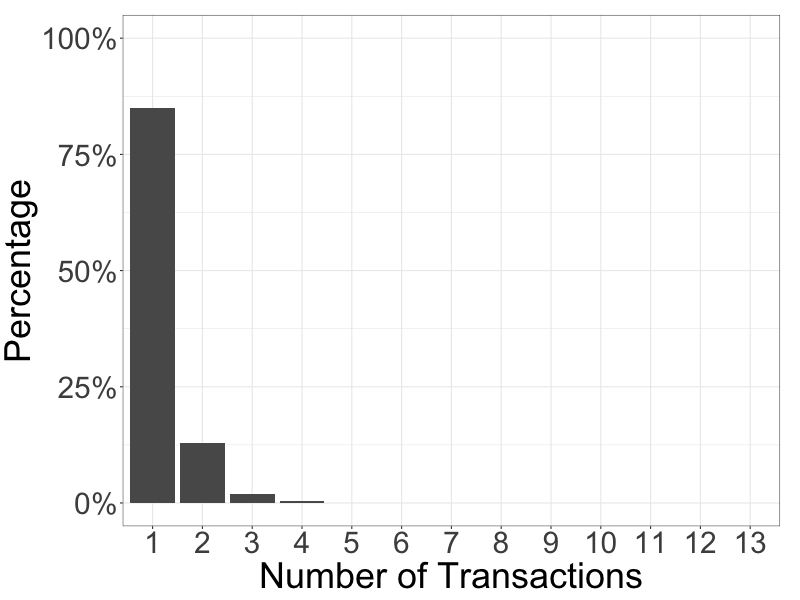}
      \caption{Aurory}
      \label{fig:9_1}
    \end{subfigure}\hfil 
    \begin{subfigure}[b]{0.4\textwidth}
      \includegraphics[width=\linewidth]{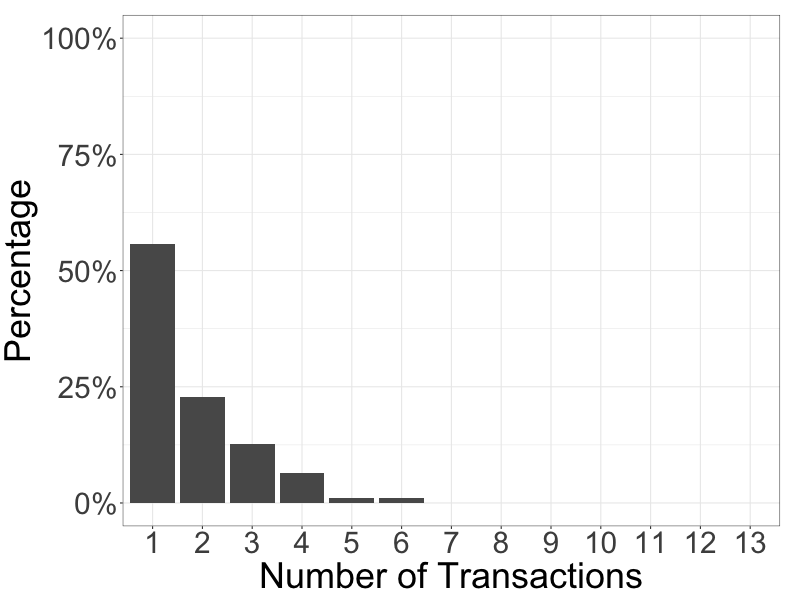}
      \caption{Autoglyphs}
      \label{fig:9_2}
    \end{subfigure}\hfil 
    \medskip
    \begin{subfigure}[b]{0.4\textwidth}
      \includegraphics[width=\linewidth]{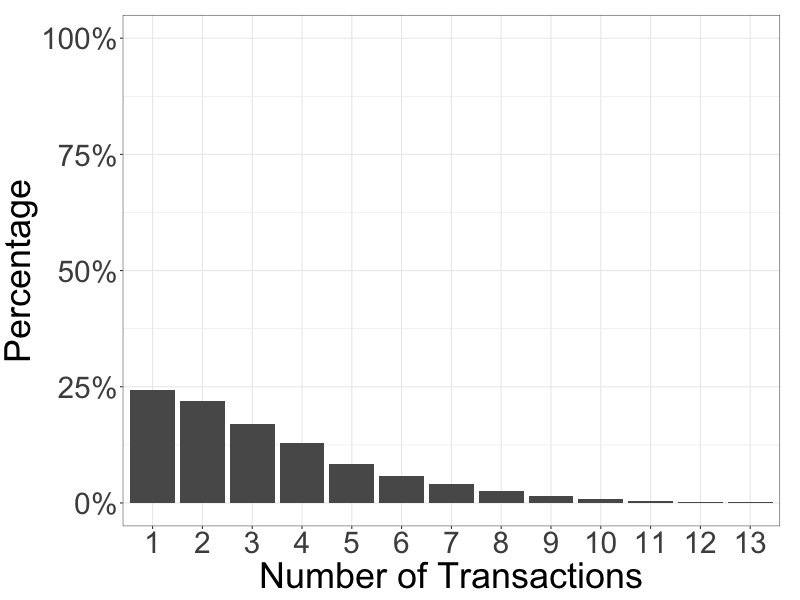}
      \caption{Bored Ape Kennel Club}
      \label{fig:9_3}
    \end{subfigure}\hfil
    \begin{subfigure}[b]{0.4\textwidth}
      \includegraphics[width=\linewidth]{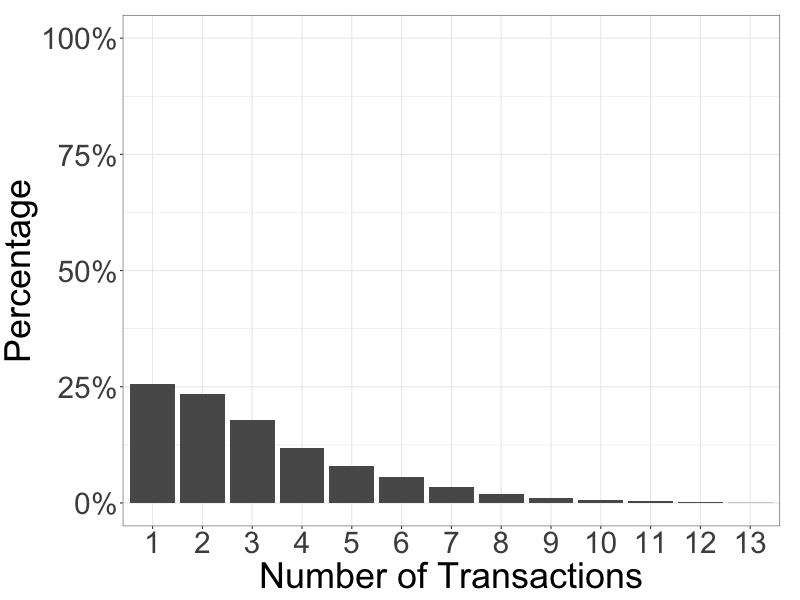}
      \caption{Bored Ape Yacht Club}
      \label{fig:9_4}
    \end{subfigure}\hfil 
    \medskip
    \begin{subfigure}[b]{0.4\textwidth}
      \includegraphics[width=\linewidth]{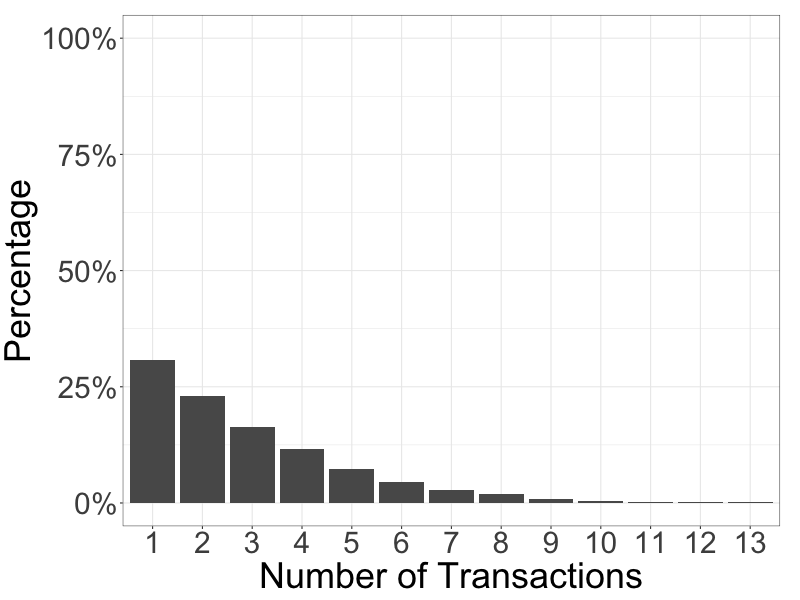}
      \caption{Cryptopunks}
      \label{fig:9_5}
    \end{subfigure}\hfil 
    \begin{subfigure}[b]{0.4\textwidth}
      \includegraphics[width=\linewidth]{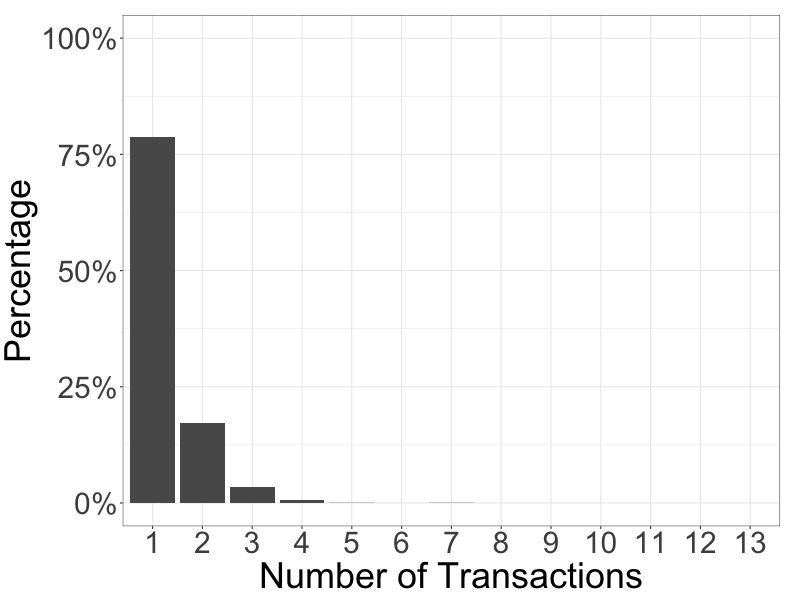}
      \caption{Degenerate Ape Academy}
      \label{fig:9_6}
    \end{subfigure}\hfill
    \medskip
    \begin{subfigure}[b]{0.4\textwidth}
      \includegraphics[width=\linewidth]{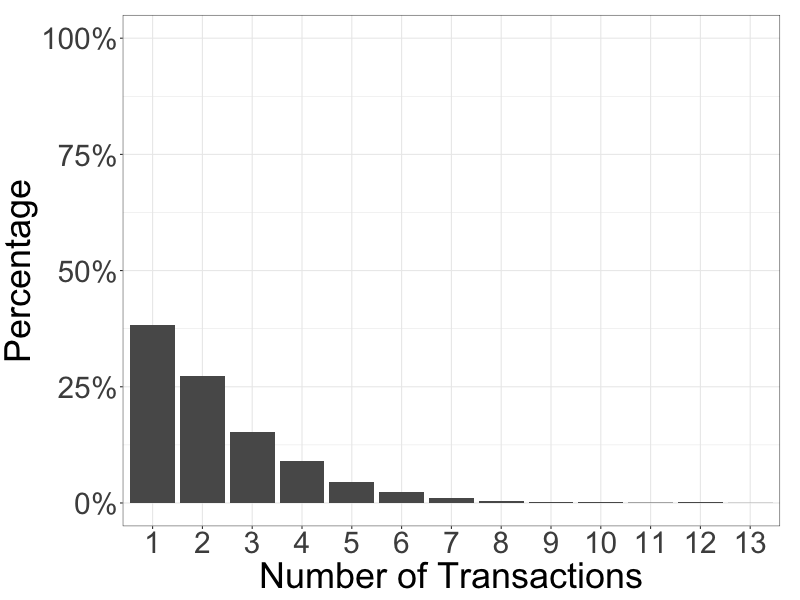}
      \caption{Meebits}
      \label{fig:9_7}
    \end{subfigure}\hfil 
    \begin{subfigure}[b]{0.4\textwidth}
      \includegraphics[width=\linewidth]{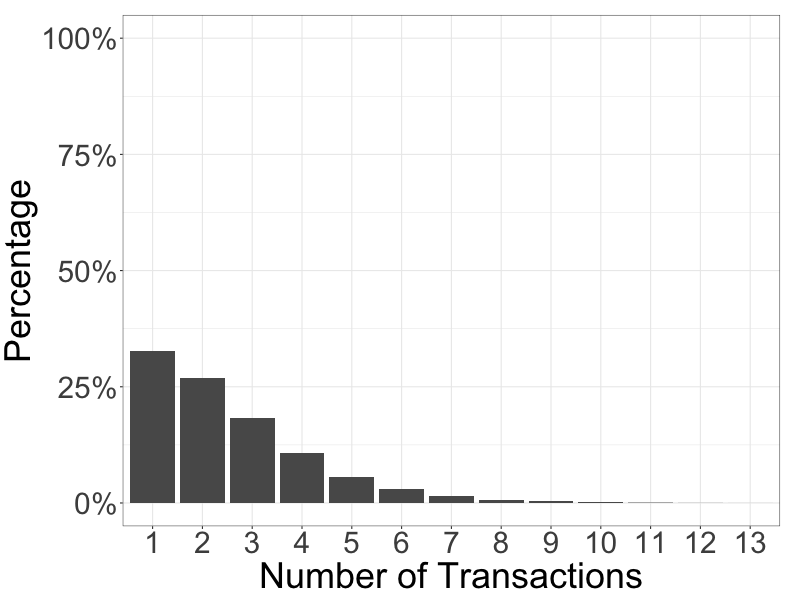}
      \caption{Mutant Ape Yacht Club}
      \label{fig:9_8}
    \end{subfigure}\hfil 
    \caption{Percentage of the Number of Transactions per Token.} \label{fig:Transaction_counts}
\end{figure}

Another challenge associated with the analysis of a NFT collections is its extreme volatility, as has been pointed out by \cite{DOWLING_PRICE}, \cite{Borri2022-ki}, \cite{KongLin} and \cite{Mazur2021-cy}. Unlike cryptocurrencies or equity markets where a large number of historical data points exist at a regular interval, the NFT market is highly illiquid with sales occurring sporadically over time as seen in Figure~\ref{fig:Transactions}. This may be one driver that causes the transacted prices to be extremely volatile, as suggested by~\cite{kapooretal}. Moreover, a large proportion of tokens are transacted only once after the collection launch. Figure~\ref{fig:Transaction_counts} shows the distribution of the number of transactions of each token during the studied period and it also suggests that only a small fraction of tokens are traded more than three times. 

\begin{figure}[h]
    \centering 
    \begin{subfigure}[b]{0.4\textwidth}
      \includegraphics[width=\linewidth]{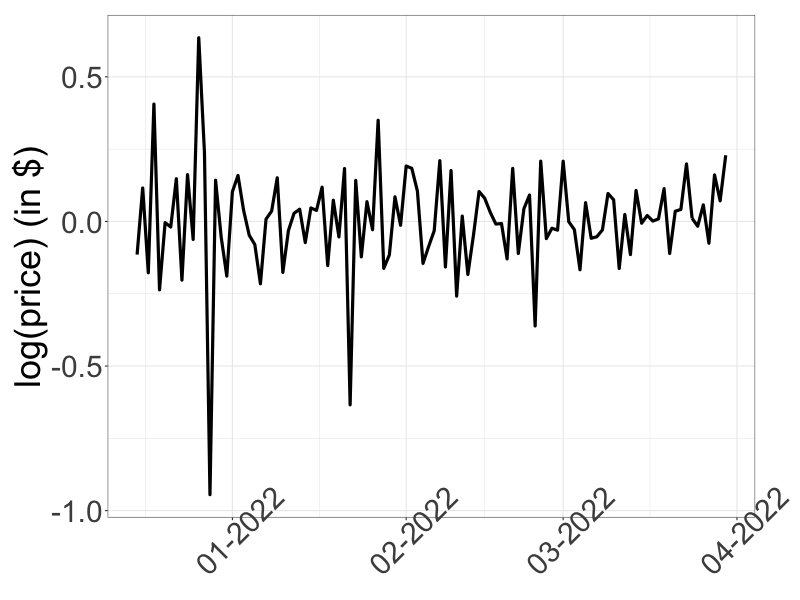}
      \caption{Aurory}
      \label{fig:10_1}
    \end{subfigure}\hfil 
    \begin{subfigure}[b]{0.4\textwidth}
      \includegraphics[width=\linewidth]{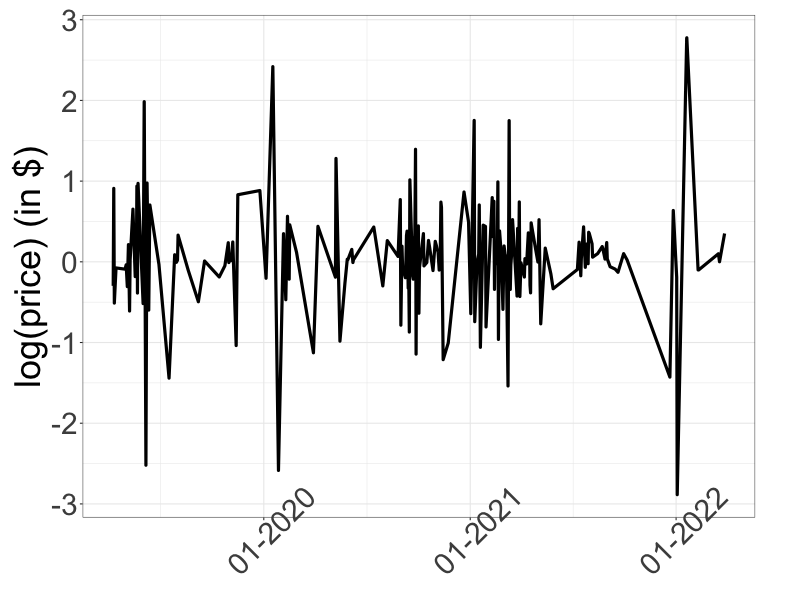}
      \caption{Autoglyphs}
      \label{fig:10_2}
     \medskip
    \end{subfigure}\hfil 
    \begin{subfigure}[b]{0.4\textwidth}
      \includegraphics[width=\linewidth]{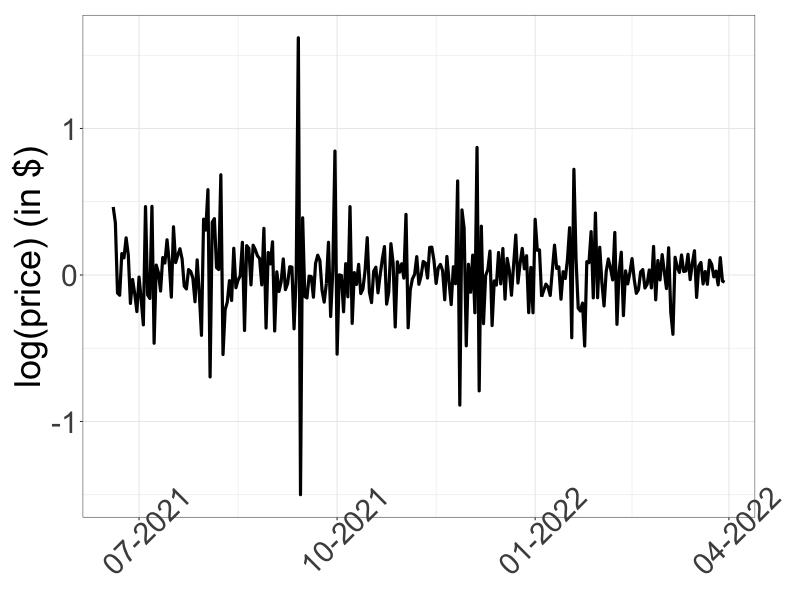}
      \caption{Bored Ape Kennel Club}
      \label{fig:10_3}
    \end{subfigure}\hfil
    \begin{subfigure}[b]{0.4\textwidth}
      \includegraphics[width=\linewidth]{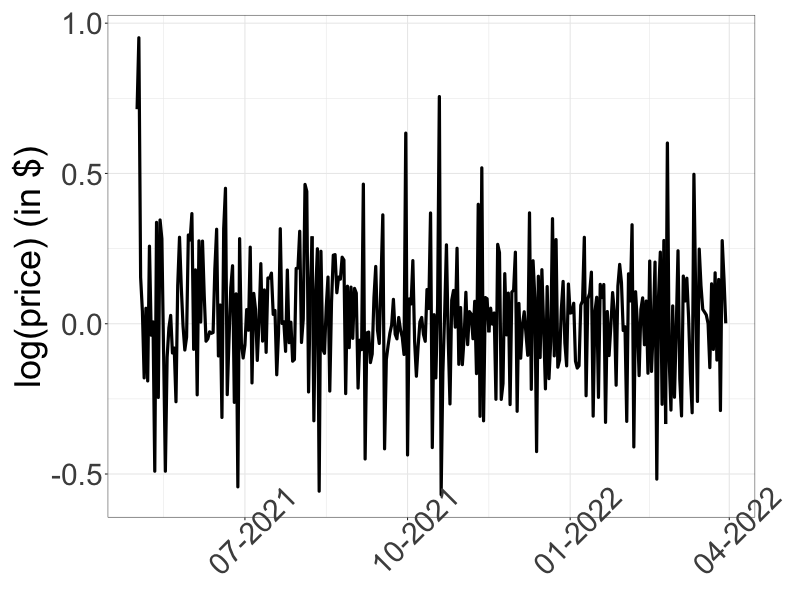}
      \caption{Bored Ape Yacht Club}
      \label{fig:10_4}
    \end{subfigure}\hfil 
    \medskip
    \begin{subfigure}[b]{0.4\textwidth}
      \includegraphics[width=\linewidth]{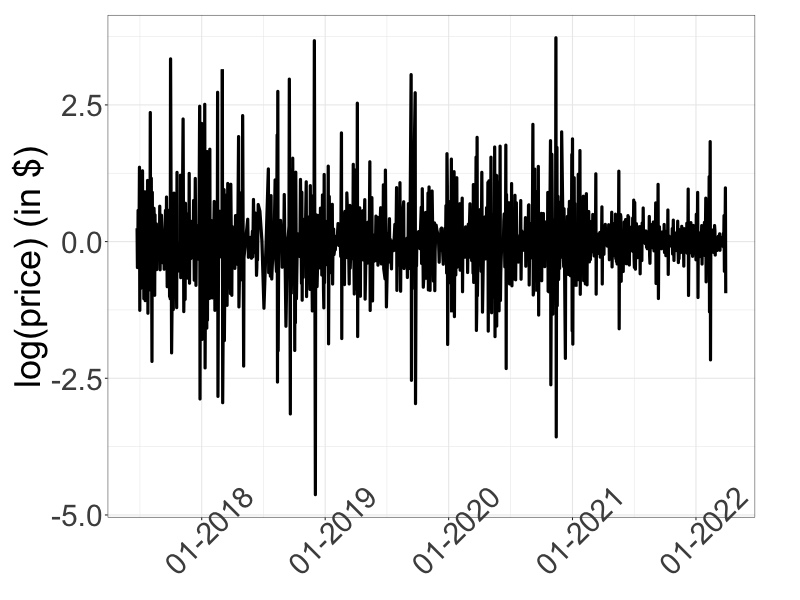}
      \caption{Cryptopunks}
      \label{fig:10_5}
    \end{subfigure}\hfil 
    \begin{subfigure}[b]{0.4\textwidth}
      \includegraphics[width=\linewidth]{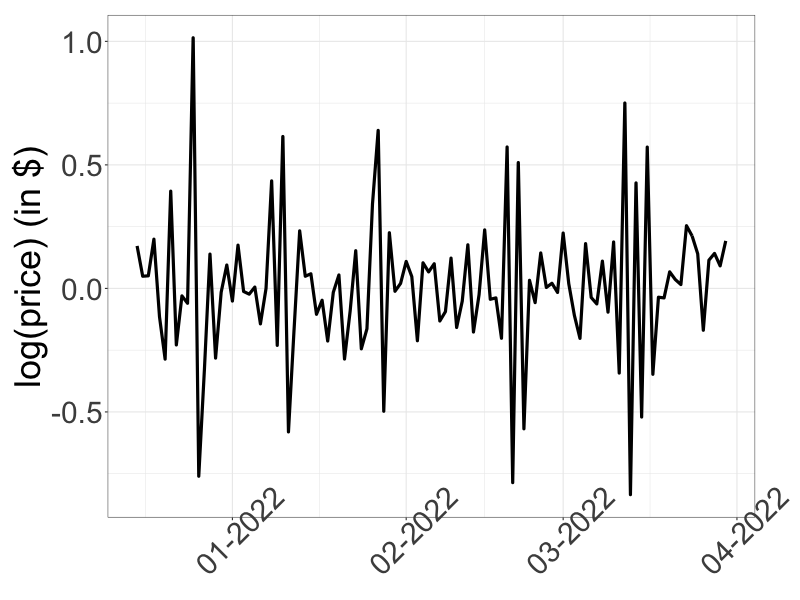}
      \caption{Degenerate Ape Academy}
      \label{fig:10_6}
    \end{subfigure}
    \medskip
    \begin{subfigure}[b]{0.4\textwidth}
      \includegraphics[width=\linewidth]{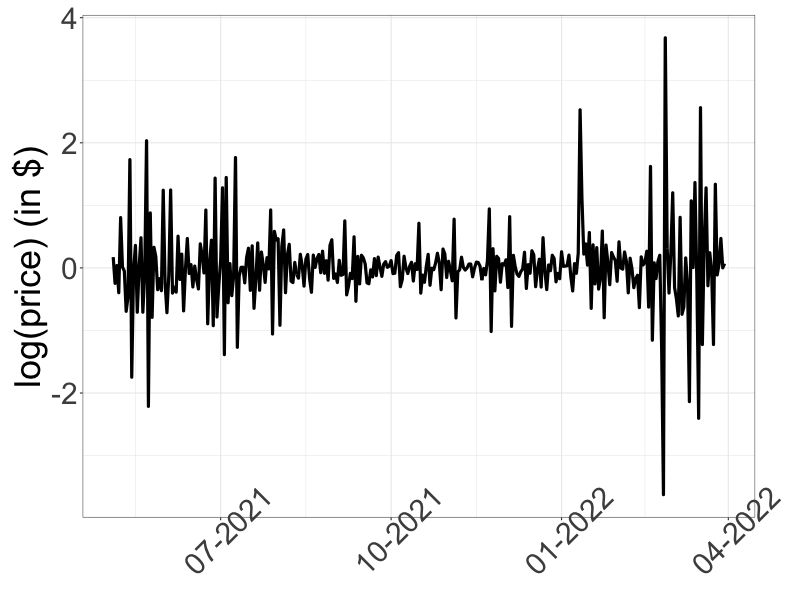}
      \caption{Meebits}
      \label{fig:10_7}
    \end{subfigure}\hfil 
    \begin{subfigure}[b]{0.4\textwidth}
      \includegraphics[width=\linewidth]{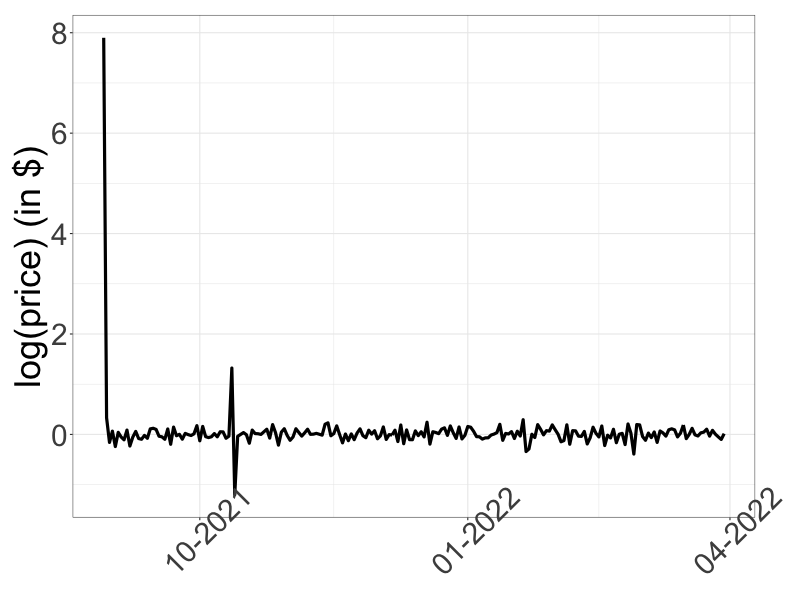}
      \caption{Mutant Ape Yacht Club}
      \label{fig:10_8}
    \end{subfigure}\hfil 
    \caption{Average Daily Log Return by Collections.} \label{fig:Log_Return}
\end{figure}

Indeed, the volatility of all eight studied collections during the studied period was extremely high (Figure~\ref{fig:Log_Return}). The average daily log returns at the collection level were rather stationary except for Mutant Ape Yacht Club. Average returns for Mutant Ape Yacht club were especially volatile for the first few days but became stationary with relatively constant variance throughout the studied period. 12-month realized volatility, the standard deviation of the daily log return times the square root of the number of trading days of 252, was computed for the eight collections based on the daily average transacted price of tokens (\ref{tab:volatility}). Cryptopunks had the highest realized volatility of 1,140\% and Aurory had the lowest with 291\%. For comparison, the annualized volatility of S\&P 500 was 14.95\% as of March 31st of 2022. The 52-week high of volatility for crude oil, a commodity that is known to be volatile, in the past year was around 102\%, according to CBOE Crude Oil Volatility Index (OVX). So far, studies have addressed this issue by either averaging the data or aggregating the data to a weekly or monthly level. For example, \cite{kapooretal} studied the relationship between the social media mentions of a given NFT token and its price. In the analysis, the token value was defined to be the average selling price of the asset over all its historic sales. \cite{DOWLING_PRICE} studied the pricing of NFT of land in a virtual world. Due to extreme volatility and the low number of daily transactions at the daily level, data were aggregated at the weekly level and were used for the analysis.

\begin{table}[h]
\centering
\resizebox{\textwidth}{!}{%
\begin{tabular}{llllll}
\cline{1-5}
\multicolumn{1}{|l|}{} & \multicolumn{1}{c}{Aurory} & \multicolumn{1}{c}{Autoglphs} & \multicolumn{1}{c}{Bored Ape Kennel Club} & \multicolumn{1}{c|}{Bored Ape Yacht Club} &  \\ \cline{1-5}
\multicolumn{1}{|l|}{12-month Realized Volatility} & \multicolumn{1}{r}{291\%} & \multicolumn{1}{r}{1,084\%} & \multicolumn{1}{r}{420\%} & \multicolumn{1}{r|}{340\%} &  \\ \cline{1-5}
 &  &  &  &  &  \\ \cline{1-5}
\multicolumn{1}{|l|}{} & \multicolumn{1}{c}{Cryptopunks} & \multicolumn{1}{c}{Degenerate Ape Academy} & \multicolumn{1}{c}{Meebits} & \multicolumn{1}{c|}{Mutant Ape Yacht Club} &  \\ \cline{1-5}
\multicolumn{1}{|l|}{12-month Realized Volatility} & \multicolumn{1}{r}{1,140\%} & \multicolumn{1}{r}{464\%} & \multicolumn{1}{r}{1,013\%} & \multicolumn{1}{r|}{896\%} &  \\ \cline{1-5}
 &  &  &  &  & 
\end{tabular}%
}
\caption{12 Month Realized Volatility of eight profile picture collections}
\label{tab:volatility}
\end{table}

\section{Conclusions and Outlook}

The emergence of NFTs initially spurred interest from blockchain enthusiasts and individual collectors. In recent years, though, as a growing number of NFT collections have become highly valued, the NFT ecosystem has {\em financialized} and NFTs can now be regarded as a novel financial asset class. Financial decisions rely on data and analysis thereof. This paper has described characteristics common to transactional data for many NFT collections and has illustrated that such data entail unique analytical challenges. Owing to these and other challenges, several authors have identified NFTs as an emerging field of research. Beyond doubt, the existing body of academic literature is still limited and many research questions regarding NFTs remain unanswered. \cite{Baals2022} recognize this fact and propose a research agenda forward with focus on the financial economics for NFTs. While research on financial economics for NFTs will encompass advanced data analytics, future research NFT transactions may offer a set of opportunities beyond that field as well, particularly in data science. 

In fact, each of the challenges described in Section \ref{sec:Challenges} may lead to novel data science methods and/or applications being proposed. For instance, to account for the high price differentiation according to certain, but not all, traits, one can imagine to construct a machine learning model that uses information from both transactions and token metadata, possibly even from the images themselves, to obtain accurate token specific price predictions. Section \ref{sec:wash} hinted to ways to detect wash trading at the transaction and collection level. These ideas could be used to construct a robust detection algorithm. Moreover, NFT data that involve wash trades can be a fertile ground for classification and anomaly detection models, such as robust statistics. At the height of the Meebits wash trade wave caused by LooksRare's reward system, more than half of transactions were suspect wash trades. A fraction of over 50\% of anomalies generally poses challenges to be detected by robust statistics as the latter will break down, which provides an interesting avenue for research. Finally, the volatility challenges highlighted in Section \ref{sec:Volatility} require deeper investigation, should one want to apply downstream financial analytics, such as risk/reward quantification or NFT portfolio optimization. 

As presented, NFT data provide a fertile ground for future data science research with its own unique challenges. We hope that the material presented here can be a seed for further developments in the field and that the data shared here prove to be useful in bringing about such developments.  

\newpage
\bibliographystyle{plainnat}
\bibliography{ref}

\end{document}